\documentclass[a4paper]{article}

\usepackage{icrc2013}
\usepackage{textcomp}
\usepackage[english]{babel}

\title{FACT - The First G-APD Cherenkov Telescope: Status and Results}

\shorttitle{FACT - The First G-APD Cherenkov Telescope: Status and Results}

\newcommand{\ethz}{$^1$}
\newcommand{\tudo}{$^2$}
\newcommand{\unige}{$^3$}
\newcommand{\uniw}{$^4$}
\newcommand{\epfl}{$^5$}

\newcommand{\uniz}{$^{a}$}
\newcommand{\unizh}{$^{^a}$}
\newcommand{\kynu}{$^{b}$}
\newcommand{\kynuh}{$^{^b}$}
\newcommand{\mpim}{$^{c}$}
\newcommand{\mpimh}{$^{^c}$}
\newcommand{\tum}{$^{d}$}
\newcommand{\tumh}{$^{^d}$}

\authors{
T.~Bretz\ethz,
H.~Anderhub\ethz,
M.~Backes\tudo,
A.~Biland\ethz,A
V.~Boccone\unige,
I.~Braun\ethz,
J.~Bu\ss\tudo,
F.~Cadoux\unige,
V.~Commichau\ethz,
L.~Djambazov\ethz,
D.~Dorner\uniw,
S.~Einecke\tudo,
D.~Eisenacher\uniw,
A.~Gendotti\ethz,
O.~Grimm\ethz,
H.~von Gunten\ethz,
C.~Haller\ethz,
D.~Hildebrand\ethz,
U.~Horisberger\ethz,
B.~Huber\ethz\unizh,
\mbox{K.-S.}~Kim\ethz\kynuh,
M.~L.~Knoetig\ethz,
\mbox{J.-H.}~K\"ohne\tudo,
T.~Kr\"ahenb\"uhl\ethz,
B.~Krumm\tudo,
M.~Lee\ethz\kynuh,
E.~Lorenz\ethz\mpimh,
W.~Lustermann\ethz,
E.~Lyard\unige,
K.~Mannheim\uniw,
M.~Meharga\unige,
K.~Meier\uniw,
T.~Montaruli\unige,
D.~Neise\tudo,
F.~Nessi-Tedaldi\ethz,
\mbox{A.-K.}~Overkemping\tudo,
A.~Paravac\uniw,
F.~Pauss\ethz,
D.~Renker\ethz\tumh,
W.~Rhode\tudo,
M.~Ribordy\epfl,
U.~R\"oser\ethz,
\mbox{J.-P.}~Stucki\ethz,
J.~Schneider\ethz,
T.~Steinbring\uniw,
F.~Temme\tudo,
J.~Thaele\tudo,
S.~Tobler\ethz,
G.~Viertel\ethz,
P.~Vogler\ethz,
R.~Walter\unige,
K.~Warda\tudo,
Q.~Weitzel\ethz,
M.~Z\"anglein\uniw $\;\;$
(FACT Collaboration)
}
\afiliations{
\ethz ETH Zurich, Switzerland $\;\;-\;\;$
   Institute for Particle Physics, Schafmattstr.~20, 8093 Zurich\\
\tudo Technische Universit\"at Dortmund, Germany $\;\;-\;\;$
   Experimental Physics 5, Otto-Hahn-Str.~4, 44221 Dortmund\\
\unige University of Geneva, Switzerland $\;\;-\;\;$
   ISDC, Chemin d'Ecogia~16, 1290 Versoix $\;\;-\;\;$
   DPNC, Quai Ernest-Ansermet 24, 1211 Geneva\\
\uniw Universit\"at W\"urzburg, Germany $\;\;-\;\;$
   Institute for Theoretical Physics and Astrophysics,
   Emil-Fischer-Str.~31, 97074 W\"urzburg\\
\epfl EPF Lausanne, Switzerland  $\;\;-\;\;$
   Laboratory for High Energy Physics, 1015 Lausanne\\
\scriptsize{
\uniz {Also at: University of Zurich, Physik-Institut, 8057 Zurich,
   Switzerland}\\
\kynu {Also at: Kyungpook National University, Center for High Energy
   Physics, 702-701 Daegu, Korea}\\
\mpim {Also at: Max-Planck-Institut f\"ur Physik, D-80805 Munich,
   Germany}\\
\tum  {Also at: Technische Universit\"at M\"unchen, D-85748 Garching,
   Germany}\\
}
}

\email{thomas.bretz@phys.ethz.ch}

\abstract{
The First G-APD Cherenkov telescope (FACT) is the first telescope using
silicon photon detectors (G-APD aka. SiPM). It is built on the mount of
the HEGRA CT3 telescope, still located at the Observatorio del Roque de
los Muchachos, and it is successfully in operation since Oct. 2011. The
use of Silicon devices promises a higher photon detection efficiency,
more robustness and higher precision than photo-multiplier tubes. The
FACT collaboration is investigating with which precision these devices
can be operated on the long-term. Currently, the telescope is
successfully operated from remote and robotic operation is under
development. During the past months of operation, the foreseen
monitoring program of the brightest known TeV blazars has been carried
out, and first physics results have been obtained including a strong
flare of Mrk501. An instantaneous flare alert system is already in a
testing phase. This presentation will give an overview of the project
and summarize its goals, status and first results.}


\begin{document}
\maketitle

\section{Introduction}

\begin{figure}[t]
 \centering
 \includegraphics[width=0.46\textwidth,angle=0,clip]{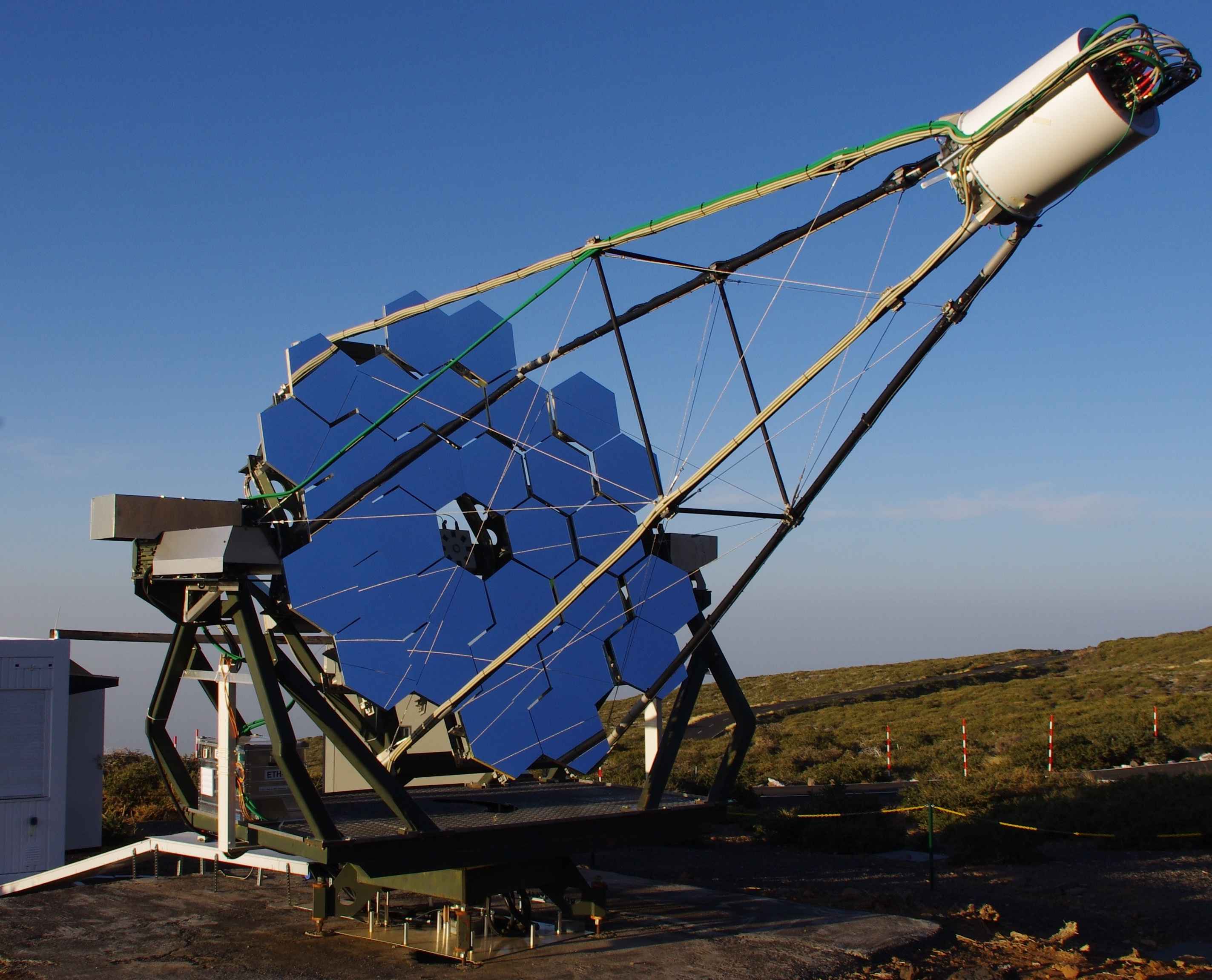}
 \caption{Picture of the telescope}
 \label{fig:photo}
\end{figure}

Since Oct.\ 2011, the FACT Collaboration is operating the First G-APD
Cherenkov telescope~\cite{bib:design} at the Observatorio del Roque de los Muchachos  on
the Canary Island of La Palma. The telescope's camera is the first
focal plane installation using Geiger-mode avalanche photo-diodes
(G-APD) for photo detection. Comprising 1440 channels individually read
out, each pixel has a field-of-view of 0.11\textdegree{} yielding
a total field-of-view of 4.5\textdegree{}. The camera is installed
on the mount of the former HEGRA CT\,3 telescope. After replacement
of the old disc shaped mirrors with refurbished hexagonal mirrors,
it has now a total reflective surface of 9.5\,m\(^2\). A picture
of the telescope is shown in Fig.~\ref{fig:photo}.

With this novel camera, silicon photo sensors have started to replace photo
multiplier tubes (PMT) still widely used for photo detection.
They offer a high gain ($10^5$ to $10^6$) and are robust enough to be operated
under moon light conditions. Their single photon counting capability,
their compactness and their low operational
voltage ($<$\,100\,V) simplifies the camera design. The photon detection
efficiency (PDE) of commercially available G-APDs is already at the
level of the best PMTs, and will still significantly improve in the
future.

The challenge of the project is to understand and operate the
silicon based photo sensors under changing environmental conditions,
such as changing auxiliary temperature or changing photon flux
from the diffuse night-sky background. A system to keep the
G-APD response stable under these conditions has been developed.

The telescope is dedicated to long-term monitoring of the brightest
known TeV blazars. These highly violent objects show variability
on time scales of seconds to years. To study their behavior 
on the longer time scales, a complete data sample obtained 
from continuous measurements on long time scales is necessary.



\section{System overview}

The camera of the telescope is compiled from 1440 channels. Each channel
is equipped with a G-APD (Hamamatsu MPPC S10362-33-50C) and a solid light
concentrator. The solid light concentrators have the advantage, as 
compared to hollow ones, to feature a better compression ratio between
entrance window and exit and thus balance the relatively small size
of the sensors (9\,mm\(^2\)). Sensors, cones and a protective window
were glued together using optical glue. All channels are individually
read out by the Domino Ring Sampling chip (DRS\,4). 
The data is transferred by a TCP/IP connection via Ethernet to a 
data acquisition PC.
The summed signal 
of nine channels form a trigger signal which is discriminated by 
a comparator. Each trigger patch is divided in two bias voltage channels
with four and five G-APDs, resp. Although all read out electronics
was integrated into the camera, the bias voltage supply is located
in the counting hut. Each bias voltage channel has its own
current readout. The precision of the voltage application is 12\,bit
up to 90\,V and 12\,bit up to 5\,mA for the current readout.

The total power consumption of the camera electronics during operation
is about 570\,W. About 100\,W are dissipated in the supply lines and
another 100\,W due to the limited efficiency of the DC-DC converters,
yielding a total of 370\,W consumed by the electronics,
\(\approx\)260\,mW per channel.

To get rid of the waste heat produced by the camera electronics,
a passive water cooling is applied. To avoid the sensors being
heat up by the electronics, a thermal insulation is installed in between.
In total, 31 temperature sensors measure the
temperature of the sensor plane.

All details about the camera construction, the electronics and control
software can be found in~\cite{bib:design}.

\section{Status and achievements}
\subsection{Operation}

\begin{figure}[h]
 \centering
 \includegraphics[width=0.46\textwidth,angle=0,clip]{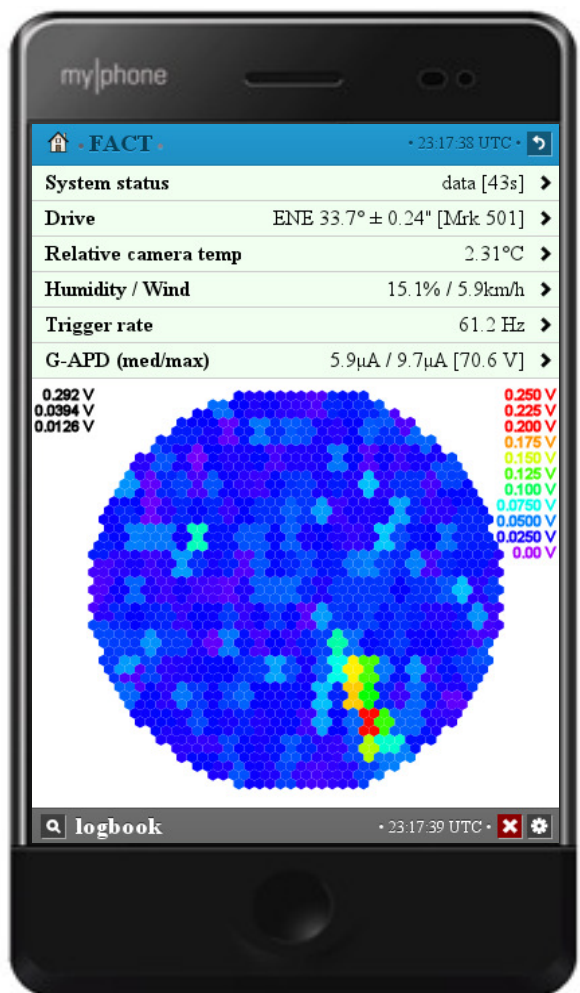}
 \caption{Web interface. To achieve low network
traffic, only the average signal is displayed for four and five pixels.
\url{http://www.fact-project.org/smartfact/}}
 \label{fig:smartfact}
\end{figure}

For a high duty cycle and consistent and stable data taking,
a robust slow control and data acquisition software is needed.
To avoid the need of shift crew on site, full remote operation
must be possible. 
The camera is now operated since 20 months and 
its auxiliary hardware has been upgraded gradually, to allow for 
full remote control. This includes an interlock system, which purpose is
to protect the camera from a cooling failure, Ethernet switchable 220\,V plugs
and the Ethernet access to all power supplies and a remote controllable lids.
More details
on that can be found in~\cite{bib:robotic}.
In parallel, a fully automatic control system
has been developed. While all important parameters of the system
are available via Web interface (see Fig.~\ref{fig:smartfact}), the control of the system
is achieved from a JavaScript engine reading the schedule from a
database. Since a couple of weeks, this system is operating, and
user interaction is usually only required in case of unexpected
weather conditions.

Currently, a fully automatic analysis is also in the testing phase. It
is applied on the data immediately after the data has been recorded and the
file has been closed which typically happens every 5\,min for physics data.
The analysis takes about the same time than data taking and will allow
prompt flare alerts for other telescopes in the near future.

\subsection{Stability}

One of the main goals of the project was the prove of the
applicability of G-APDs during standard operation conditions.
Since the gain and all other properties of G-APDs depend on their
temperature and the voltage applied, a correcting element is 
necessary. By design, the temperature gradient in the focal plane
is small enough allowing several G-APDs to be connected to a single bias voltage 
channel. In total, 31 temperature sensors installed in the focal plane
allow to counteract temperature changes by adapting the applied bias voltage accordingly.
The power supply of each sensor is equipped with a filter network of resistors.
Due to the changing light conditions induced by changes of the
atmospheric conditions or moon light, the current drawn by the
G-APDs is varying inducing a voltage drop in the resistances.
To correct for this voltage drop, the current of each channel is measured
and the voltage corrected accordingly.

To measure the stability of the gain with these two feedback loops applied,
three different methods are available:
\paragraph{External light pulser} An external temperature stabilized
light pulser triggers the camera. Since its light yield is stable
on average, the gain can directly be deduced from the average pulse height.
However, the precision of this system measuring the gain of the G-APD
is limited by the fact that also the transmission of the entrance window
and the cones could change with time.
\paragraph{Dark count spectrum} A direct and mostly unbiased method to
extract the gain, is the
extraction of the dark count spectrum. It only depends on the performance
of the G-APDs and the readout chain. Its drawback is that it can only be determined
under light conditions, i.e. count rates, which still allow for extraction
of single pulses. This is the case with closed lids or at dark time conditions.
\paragraph{Ratescans} To measure the stability of the gain in moon lit nights,
so called ratescans are used. Ratescans
measure the trigger response as a function of the applied trigger threshold.
While at low thresholds, the very high rates are dominated by random coincidences 
of background photons, at high thresholds, the rate represents 
the number of triggers from coincident light flashed, induced mainly
by hadronic showers. Although this method
depends on the performance of the whole system, it allows to measure the
response of the system even at the brightest light conditions.
To ensure an unbiased measurement, only ratescans taken during good
atmospheric conditions and weather conditions must be taken into account.
A more detailed discussion of this can be found in~\cite{bib:ratescans}.\\

All three methods have been applied almost daily since the beginning of
operation. This allows to compare the response at any possible
light condition and all temperatures occurring over the seasons.
All methods have shown consistently that with the applied feedback
system, a stable gain up to full moon conditions can be achieved.
From the dark count spectrum, a stability of better than \(\pm\)3\% over time
and 4\% pixel-to-pixel variations was deduced.
Currently, the limiting factor of the system is the calibration of the
applied bias voltage which is under investigation and will be improved soon.

More details about the stability can be found in~\cite{bib:stability} and
will be available soon in~\cite{bib:feedback}.

\subsection{Data acquisition}
The Ethernet connection between the forty readout boards and the
data acquisition PC is routed over four Ethernet switches, two located in the
camera and two located in the control room. In total, four Ethernet lines
are available, one Ethernet line per ten boards. A total maximum transfer
rate of about 1.9\,Gbps is achieved. This is limited by the maximum
throughput of 50\,Mbps of each of the Ethernet-chips (Wiznet 5300) on-board of the
forty data acquisition boards.
The transfer rate corresponds to a trigger
rate of \(\sim\)250\,Hz for a readout window of 300 samples and \(\sim\)80\,Hz 
for the readout of the full DRS pipeline (1024 samples).

\subsection{Trigger threshold setting}
One of the most important aspects during data taking is the correct setting
of the trigger thresholds. To avoid high rates from random coincidences of
background photons, the need to be set above the noise level. Since the
noise level, mainly defined by the diffuse light from the night sky background,
is not necessarily constant, some contingency is added to avoid strong rate
fluctuations during data taking. At the same
time, they must be set to as low as possible to achieve the lowest energy 
threshold and thus the highest integral sensitivity. 
Since the number of breakdowns in the G-APD, i.e. the number of detected 
photons, is proportional to the measured current, the rate induced
from random coincidences as a function of the trigger threshold can 
directly be determined from the current.
From ratescans taken at different light conditions over more
than a year, a parametrization for the
rate induced from background light and for the rate induced by hadronic showers
under best conditions, was derived. Since several weeks, the trigger threshold
of the system is now directly deduced from the measured current resulting in
very stable data taking conditions. To allow for easy analysis, it is kept constant
during every 5\,min run.

\subsection{Current and energy threshold prediction}
To allow for an efficient schedule, the prediction of the system response
under changing light conditions is mandatory. Therefore, the measured
current has been correlated with moon conditions, i.e. moon brightness and
zenith angle, which leads to a very precise prediction of the measured
current. Since the trigger threshold is in the first order proportional
to the light yield of the triggered shower, the change in energy threshold
can directly be deduced. Including the change of the light yield with
zenith angle, gives a very good estimate of the raise in energy threshold,
i.e. loss in sensitivity, for each observation.

These topics are discussed in more details in~\cite{bib:threshold,bib:moon}
and will be available soon in~\cite{bib:feedback}.

\section{Physics results}

Very soon after the first observations, first results could be
presented~\cite{bib:Gamma2012}. From the measurements of the Crab Nebula,
a sensitivity of 8\% Crab in 50\,h could be derived. This
is based on the analysis of data taking during the first months of operation 
and based on an analysis primarily developed for MAGIC not yet
necessarily optimized for the current system. From the rate of excess events
an energy threshold between 400\,GeV and 700\,GeV can be derived 
taking the known spectrum of the source into account

The analysis, especially the automatic analysis, and first results
are discussed in~\cite{bib:physics}.

\begin{figure}[tb]
 \centering
 \includegraphics[width=0.46\textwidth,angle=0,clip]{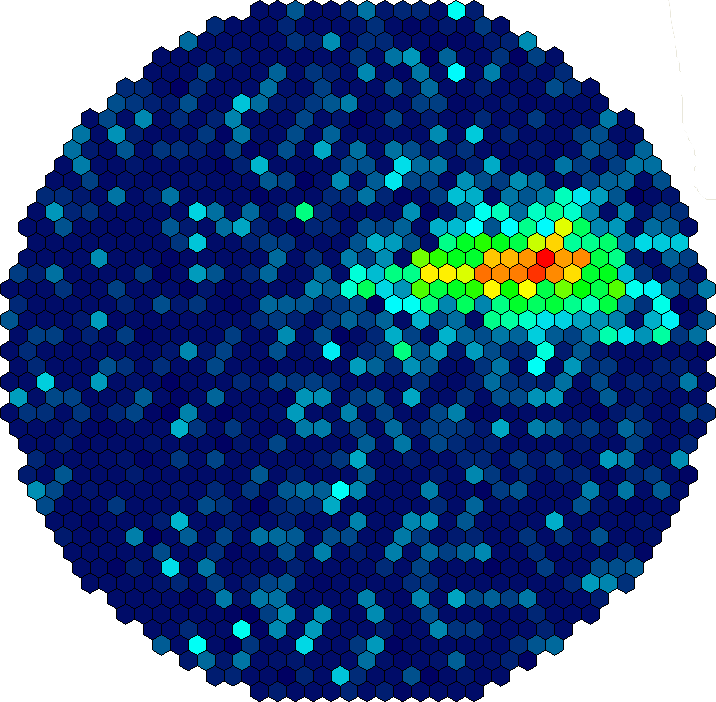}
 \caption{An exemplary event. The color scale is proportional to the signal
amplitude detected in each pixel.}
 \label{fig:event}
\end{figure}

\section{Conclusions}

The camera of the First G-APD Cherenkov Telescope (FACT) has proven the
applicability of silicon photo sensors (G-APD) in focal plane
installations, in particular as detectors in Cherenkov telescopes.
The camera is now in operation since more than one and a half years,
and although there were hardware problems during that time, none
of the was related to the G-APDs at all. So far, no decrease in
performance could be detected. Neither any hint for aging of the
G-APDs nor of the transmission of cones and window has been found.

With different types of measurements, it was shown that a
stability to on the few percent level, independent of temperature
and light conditions can be achieved. Further improvements are expected
by an improvement of the bias voltage calibration.

An important result derived from the current study is that the application
of a feedback system which keeps the G-APD's bias voltage constant,
renders the need for an external calibration device obsolete. The combination
of the extraction of the dark count spectrum and the current measurement,
is enough for the operation of such a device.

Recently, Hamamatsu presented a new generation of G-APD sensors with
significantly lower dark count rates, significantly reduced optical crosstalk
and significantly reduced afterpulse probability, expected to be available 
on the market within the next weeks~\cite{bib:hamamatsu}.

Although, it was shown that
these improvements are not obligatory for the operation in a Cherenkov telescope,
especially the reduction in optical crosstalk guarantees that the influence on 
image reconstruction is negligible.

Not only, can the FACT telescope serve as an ideal instrument for long-term
monitoring of bright blazars, the application of G-APDs in focal plane
installations will significantly reduce construction and installation costs
and improve stability of operation. This will yield a significantly impact
on future projects like The Cherenkov Telescope Array (CTA) and makes
small telescopes for monitoring purposes affordable.


\footnotesize{\paragraph{Acknowledgment}{
The important contributions from ETH Zurich grants ETH-10.08-2 and
ETH-27.12-1 as well as the funding by the German BMBF (Verbundforschung
Astro- und Astroteilchenphysik) are gratefully acknowledged.
We thank the
Instituto de Astrofisica de Canarias allowing us to operate the telescope
at the Observatorio Roque de los Muchachos in La Palma, the
Max-Planck-Institut f\"ur Physik for providing us with the mount of the
former HEGRA CT\,3 telescope, and the MAGIC collaboration for their support.
We also thank the group of Marinella Tose
from the College of Engineering and Technology at Western Mindanao
State University, Philippines, for providing us with the scheduling
web-interface.}}

\end{document}